\documentclass[conference]{IEEEtran}
\IEEEoverridecommandlockouts
\usepackage{cite}
\usepackage{amsmath,amssymb,amsfonts}
\usepackage{algorithmic}
\usepackage{graphicx}
\usepackage{textcomp}
\usepackage{xcolor}
\usepackage{subcaption}
\usepackage{float}
\usepackage{booktabs}
\usepackage[font=small]{caption}  
\usepackage{balance}
\usepackage[labelformat=simple]{subcaption}
\usepackage{array}
\usepackage{dblfloatfix} 

\def\BibTeX{{\rm B\kern-.05em{\sc i\kern-.025em b}\kern-.08em
    T\kern-.1667em\lower.7ex\hbox{E}\kern-.125emX}}
\begin{document}

\title{A 16.28 ppm/\textdegree C Temperature Coefficient, 0.5V Low-Voltage CMOS Voltage Reference with  Curvature Compensation\\
}

\author{\IEEEauthorblockN{Harshith Reddy}
\IEEEauthorblockA{\textit{Dept. of Electrical and Electronics Engineering} \\
\textit{Birla Institute of Technology and Science}\\
Pilani, India \\
f20220025@pilani.bits-pilani.ac.in}
\and
\IEEEauthorblockN{Pankaj Arora}
\IEEEauthorblockA{\textit{Dept. of Electrical and Electronics Engineering} \\
\textit{Birla Institute of Technology and Science}\\
Pilani, India \\
pankaj.arora@pilani.bits-pilani.ac.in}
}
\maketitle

\begin{abstract}
This paper presents a fully-integrated CMOS voltage reference designed in a 90 nm process node using low voltage threshold (LVT) transistor models. The voltage reference leverages subthreshold operation and near-weak inversion characteristics, backed by an all-region MOSFET model. The proposed design achieves a very low operating supply voltage of 0.5 V and a remarkably low temperature coefficient of 16.28 ppm/\textdegree C through the mutual compensation of CTAT, PTAT, and curvature-correction currents, over a wide range from -40 \textdegree C to 130 \textdegree C. A stable reference voltage of 205 mV is generated with a line sensitivity of 1.65 \%/V and a power supply rejection ratio (PSRR) of -50 dB at 10 kHz. The circuit achieves all these parameters while maintaining a good power efficiency, consuming only 0.67 \(\mu\)W.
\end{abstract}

\begin{IEEEkeywords}
CMOS voltage reference, temperature compensation, Subthreshold, power efficient.
\end{IEEEkeywords}

\section{Introduction}
In most system-on-a-chip (SoCs) systems, a voltage reference insensitive to temperature, supply voltage, process, and mismatch variations is crucial \cite{ctat-main}. They are essential components in multiple analog circuit applications such as Analog-to-Digital Converters (ADCs), linear regulators, power management systems, sensors, Internet of Things (IoT), and wearable devices \cite{21nw}. Most voltage references work by generating two types of currents, namely PTAT (proportional to absolute temperature) and CTAT (complementary to absolute temperature), and summing them to develop a temperature invariant voltage at the output \cite{ADC}.
The most used bandgap voltage reference circuits use lateral BJTs or parasitic BJTs in CMOS process, and op-amps.
In a conventional bandgap voltage reference, the base-emitter voltage (\(V_{BE}\)) of a bipolar transistor decreases linearly with temperature, exhibiting a CTAT characteristic. A PTAT voltage is generated by the difference in \(V_{BE}\) between two bipolar transistors operating at different current densities. A bandgap reference circuit combines these CTAT and PTAT voltages to produce a stable reference voltage (\(V_{REF}\)) close to the silicon bandgap voltage of 1.2 V \cite{survey}.
As CMOS technology advances, ADCs and other circuits require lower operating voltages and power-efficient performance. The start-up voltage of low-power ADCs is constrained by the minimum voltage needed for the voltage reference operation. Lowering its supply voltage extends ADC runtime and allows operation with harvested voltages as low as 0.5 V \cite{ADC}.

This work presents a purely CMOS-based voltage reference architecture that achieves an excellent temperature coefficient, with a good line sensitivity and a distinctively low operating supply voltage while maintaining power efficiency. Being purely CMOS based makes this design better compatible with a larger set of process technologies and helps achieve operating voltage by the use of low threshold transistor models and subthreshold operation. This paper is divided into four sections. Section II analyzes in detail the working of the proposed architecture. This section is subdivided into four parts to describe the different segments of the circuit – PTAT, CTAT, curvature compensation, and output reference generator. Section III discusses the simulation results achieved in this paper. Concluding remarks are made in Section IV.

\section{Proposed Design}
\begin{figure*}[htbp]
    \centering
    \includegraphics[width=\linewidth]{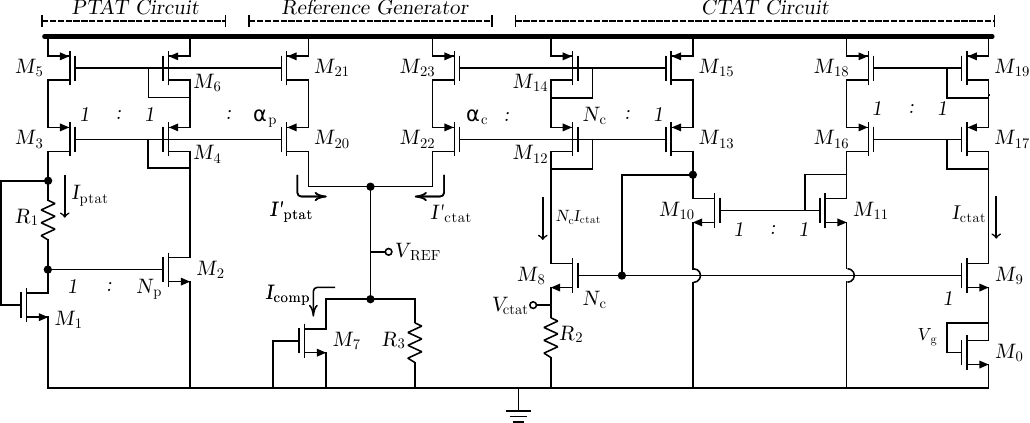}
    \caption{Schematic diagram of the proposed voltage reference circuit}
    \label{fig:full-schematic}
\end{figure*}
\subsection{PTAT Circuit}

\begin{figure}
    \centering
    \includegraphics[width=1\linewidth]{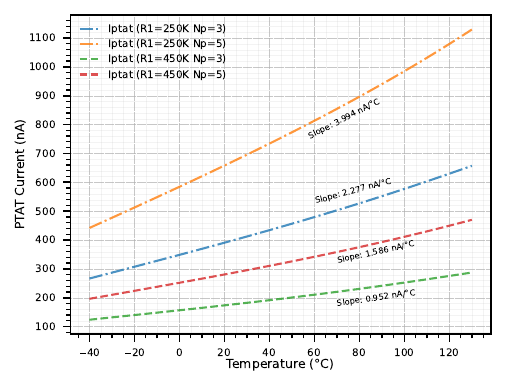}
    \caption{Drain currents of \(M_1\) and \(M_2\) (PTAT) for different \(R_1 \text{ and }N_p\), vs. temperature}
    \label{fig:ptat-current}
\end{figure}

The core part of the PTAT circuit as shown in Fig. \ref{fig:full-schematic} that generates the PTAT current are two MOSFETs, \(M_1\) and \(M_2\), operating in subthreshold region, having a similar I-V characteristics to that of a BJT, and also because of which power efficiency is significantly improved. The aspect ratios of \(M_1\) and \(M_2\) are such that, \(M_2\) is \(N_p\) times larger than \(M_1\) as marked in the schematic Fig. \ref{fig:full-schematic}, and the leakage currents, that can be defined by equation (\ref{eq:subthreshold_current}), of transistors \(M_1\) and \(M_2\) are set equal by the cascode current mirrors \(M_3\), \(M_4\), \(M_5\), and \(M_6\). Cascode current mirror helps in improving the line sensitivity of the voltage reference and reduces the voltage drop across MOSFETs \(M_1\) and \(M_2\), for them to persist in the subthreshold region across large temperature and power supply variations. The resistor \(R_1\) is necessary for maintaining a voltage drop between the gate voltages of \(M_1\) and \(M_2\) as a function of temperature. The subthreshold leakage current \(I_{ds}\) can be given as \cite{textbook}
\begin{equation}
    I_{ds} = I_{ds0} e^{\frac{V_{gs} - V_{th}}{n v_T}} 
    \left( 1 - e^{-\frac{V_{ds}}{v_T}} \right)
    \label{eq:subthreshold_current}
\end{equation}
where \(n\) is a process dependent factor caused by the depletion region characteristics and is typically in the range of 1.3 – 1.7 \cite{textbook}. The gate to source voltage is indicated as \(V_{gs}\), \(V_{ds}\) is the drain to source voltage, and \(V_{th}\) is the threshold voltage. The last term, i.e. \(( 1 - e^{-\frac{V_{ds}}{v_T}} )\), can be approximated to 1 as the \(V_{ds}\) of \(M_1\) and \(M_2\) will be biased at a voltage of at least 150 mV which is a few multiples greater than \(v_T\) even at higher temperatures (34.6 mV at 130\textdegree C). The \(\eta V_{ds}\) term in the equation 2.42 from \cite{textbook} caused due to drain-induced barrier lowering has been neglected to make calculations simpler, which can later be curvature compensated including other non-linearities. The coefficient \(I_{ds0}\) is given as \cite{textbook}

\begin{equation}
I_{ds0} = u_nC_{ox} \frac{W}{L} v_T^2e^{1.8}
    \label{eq:subthreshold_coeff}
\end{equation}
\(\mu_n\), \(C_{ox}\), \(v_T\) and \(\frac{W}{L}\) are as usual the electron mobility, per unit area gate oxide capacitance, thermal voltage and the aspect ratio of the MOSFET respectively. Now, as both the drain currents of \(M_1\) and \(M_2\) are set equal, we can say
\begin{equation}
\frac{I_{ds0,2}}{I_{ds0,1}} = e^{\frac{V_{gs1} - V_{gs2} - (V_{th1} - V_{th2})}{n v_T}}
    \label{eq:ptat-exp}
\end{equation}
If we can assume that the two threshold voltages are approximately equal (although practically they would be slightly off, which would just result in an additional constant close to 1 multiplied to our final expression), the numerator in the exponent will just result in \(V_{gs1} - V_{gs2} = I_{ds1}R_1\). Take natural log on both sides in equation (\ref{eq:ptat-exp}) we get
\begin{equation}
ln\left[\frac{(\frac{W}{L})_2}{(\frac{W}{L})_1}\right] = \frac{I_{ptat}R_1}{nv_T}
    \label{eq:ptat-ln}
\end{equation}
if we can call \(I_{ds1} = I_{ds2} = I_{ptat}\). As discussed earlier, if
\((\frac{W}{L})_2\) : \((\frac{W}{L})_1\) :: \(N_p\) : \(1\), the final expression for \(I_{ptat}\) can be derived as
\begin{equation}
I_{ptat} = \frac{nv_T}{R_1}lnN_p
    \label{eq:ptat-final}
\end{equation}
Hence we can say that the leakage currents of \(M_1\) and \(M_2\) have a PTAT nature, whose strength and the slope can be adjusted by altering \(N_p\) or the ratio of the sizes of \(M_1\) and \(M_2\), and \(R_1\) as can be seen in Fig. \ref{fig:ptat-current}.

\subsection{CTAT Circuit}\label{AA}
The CTAT architecture, as shown in Fig \ref{fig:full-schematic}, is a modified version of the voltage reference design proposed by F. Olivera and A. Petraglia \cite{ctat-main}, which exploits the multi-threshold characteristics of a MOS transistor. The unified current control model (UICM) accurately describes transistor behavior across all inversion levels that the conventional square law based models are unable to predict \cite{ctat-main}.
The core part of the CTAT circuit is the nature of the gate voltage of the MOS transistor \(M_0\), which is copied across the resistor \(R_2\) (labeled \(V_{ctat}\)) with \(N_c\) times higher current flowing in that branch. The transistors \(M_8:M_9\) and \(M_{12},M_{14}:M_{13},M_{15}\) are sized in the ratio \(N_c:1\) and, \(M_{10}:M_{11}\) and \(M_{16},M_{18}:M_{17},M_{19}\) are equally sized pairs to realize a CTAT current generator from a CTAT voltage.

\begin{figure}
    \centering
    \begin{subfigure}{\linewidth}
        \centering
        \includegraphics[width=1\linewidth]{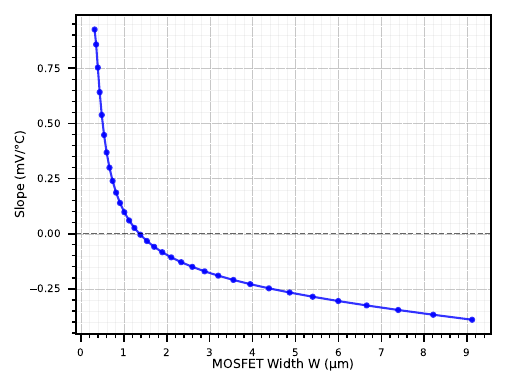}
        \caption{}
        \label{fig:ctat-slope}
    
    \medskip
    
    \end{subfigure}    \begin{subfigure}{\linewidth}
        \centering
        \includegraphics[width=1\linewidth]{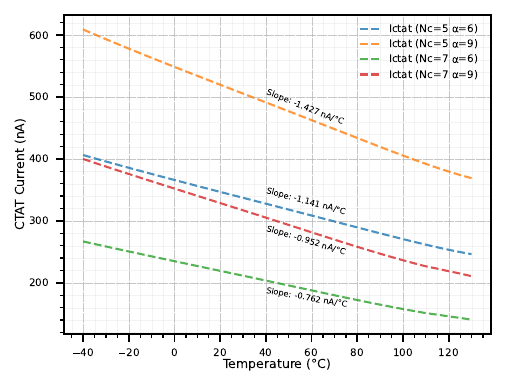}
        \caption{}
        \label{fig:ctat-current}
    \end{subfigure}
    
    \caption{CTAT Current characteristics (a) slope of CTAT gate voltage of \(M_0\) vs. width of \(M_0\) with L=10\(\mu\)m (b) CTAT current generated at output branch for different \(\alpha_c \text{ and }N_c\), vs. temperature}
    \label{fig:ctat-char}
\end{figure}

The \(M_0\) MOSFET, being diode connected, will operate mostly in forward saturation, and its drain current \(I_D\) can be given as \cite{alr-thr-appl}
\begin{equation}
I_D = I_Si_f \Rightarrow i_f = \frac{I_D}{I_S}
    \label{eq:ctat-idis}
\end{equation}
where \(i_f\) is the forward inversion level. Weak inversion is indicated by \(i_f\) $<$ 1, and similarly 1 $<$ \(i_f\) $<$ 100 and \(i_f\) $>$ 100 for moderate and strong inversion. \(I_S\) is the specific current defined by \cite{alr}
\begin{equation}
I_S = 2n \mu_n C_{ox} \frac{W}{L} v_T^2
    \label{eq:specific-current}
\end{equation}
and \(n\) is the slope factor. The I-V relationship using UICM can be modeled by \cite{alr-thr-appl}
\begin{equation}
    \begin{split}
        V_P-V_S &= v_T[\sqrt{1+i_f}  - 2 + ln(\sqrt{1+i_f} - 1)] \\
        &= v_T F(i_f)
    \end{split}
    \label{eq:all-mosfet}
\end{equation}
where, \(Vp\) the pinch-off voltage is
\begin{equation}
V_P=\frac{(V_G-V_{th0})}{n}
    \label{eq:pinch-off}
\end{equation}
\(V_S\) and \(V_G\) are the source and gate voltages w.r.t bulk (gnd), and \(V_{th0}\) is the threshold voltage at \(V_S = 0\). Equation (\ref{eq:all-mosfet}) can be re-written as
\begin{equation}
V_G = V_{th0} + n\frac{k_B}{q}F(i_f)\cdot T
    \label{eq:ctat-vg}
\end{equation}
where \(k_B\), \(q\), and \(T\) are the Boltzmann constant, electron charge, and absolute temperature, respectively. The first term being the threshold voltage of the MOS transistor expresses a CTAT behavior of approximately -0.43 mV/\textdegree C. The sign and magnitude of the function \(F(i_f)\) determines the slope of the CTAT voltage across \(M_0\), which is replicated across \(R_2\), with \(N_c\)-fold higher current as compared to \(M_0\) drain current. Hence, we can say that \(V_{ctat} = R_2N_cI_{ctat}\).
By solving this equation for \(W/L\) of \(M_0\), we get \cite{ctat-main}
\begin{equation}
\left(\frac{W}{L}\right)_{M_0} = \frac{1}{2N_cR_2}\frac{V_{th0} + n \frac{k_B}{q} F(i_f) \cdot T}{n \mu_nC_{ox} \frac{k_B^2T^2}{q^2}i_f}
    \label{eq:ctat-wl}
\end{equation}
Thus, varying the aspect ratio of \(M_0\) affects the inversion level \(i_f\) and the function \(F(i_f)\), giving the flexibility to alter the CTAT voltage slope, as shown in Fig. \ref{fig:ctat-slope}. In this design, a trade-off between linearity and power efficiency has to be done, but because of the presence of curvature compensation, a lower inversion level can be preferred to reduce power consumption. Additionally, the CTAT current ratio mirrored to the output branch can also be altered by adjusting the value of \(\alpha_c\) and \(N_c\), following the proportionality \(I_{ctat}' \propto \alpha_c \text{ and } I_{ctat}' \propto \frac{1}{N_c}\), as also shown in Fig. \ref{fig:ctat-current}.
\subsection{Curvature Compensation}
\begin{figure}
    \centering
    \includegraphics[width=1\linewidth]{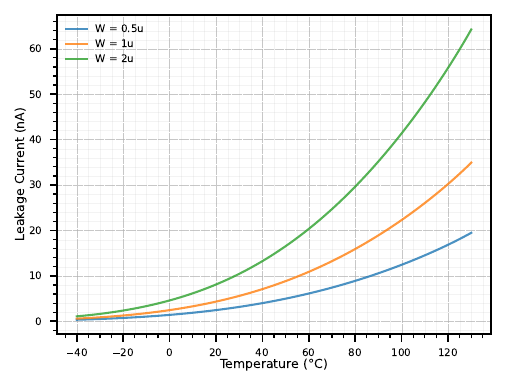}
    \caption{Subthreshold leakage current of \(M_7\) vs. temperature}
    \label{fig:leak}
\end{figure}

To better improve the temperature coefficient performance and extend the operational temperature range, curvature compensation is a widely used technique to reduce the deviation of the output voltage caused due to non-linear effects. The technique used in this design exploits the characteristics of subthreshold leakage current, which has an exponential-like property w.r.t. temperature as can be seen in Fig. \ref{fig:leak} because of the CTAT nature of the threshold voltage in the numerator of the exponent term in equation (\ref{eq:leak_current}). The strength of the curvature compensation can easily be adjusted by varying the aspect ratio of \(M_7\). The subthreshold leakage current can be given by \cite{subthreshold}
\begin{equation}
I_{Leak} = \mu_n C_{ox} \frac{W}{L} (\eta - 1) \left( \frac{k_BT}{q} \right)^2 \exp \left( -\frac{qV_{th}}{\eta k_BT} \right)
\label{eq:leak_current}
\end{equation}
where \(\eta\) is the subthreshold swing coefficient. The \(M_7\) transistor is placed in parallel to \(R_3\) such that it drains out current exponentially as temperature increases, thereby decreasing the output reference voltage across \(R_3\) and reducing the overall temperature deviation. By carefully choosing the strengths of PTAT, CTAT, and curvature compensation, an excellent temperature coefficient can be achieved.
\subsection{Output Reference Generator}
The output voltage generator circuit, as shown in Fig. \ref{fig:full-schematic}, is realized using cascode current mirrors to enhance its current mirroring capabilities - to improve line sensitivity (i.e., reference voltage variation w.r.t. \(V_{DD}\) supply) and power supply rejection \cite{psrr}. The PTAT and CTAT currents are mirrored and summed to the output branch by ratios of \(\alpha_p\) and \(\alpha_c\) respectively, through the resistor \(R_3\), generating a voltage that can be given by
\begin{equation}
V_{REF} = R_3(\alpha_pI_{ptat} + \alpha_c I_{ctat} - I_{comp})
    \label{eq:output-voltage}
\end{equation}
The value of \(V_{REF}\) is not fixed and variable by adjusting the values of \(R_3\) and current mirror ratios. The proposed circuit was designed to produce a reference voltage of 205 mV.

\section{Results Obtained}
\begin{figure}
    \centering
    \includegraphics[width=1\linewidth]{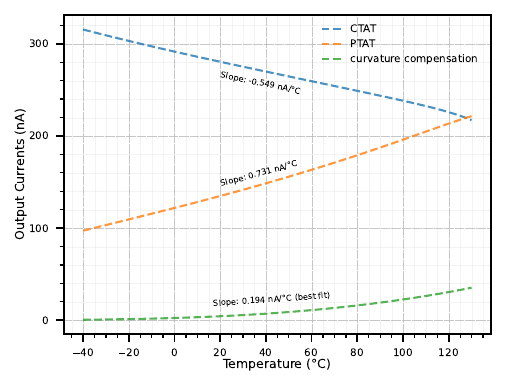}
    \caption{PTAT, CTAT and curvature compensation currents}
    \label{fig:all-currents}
\end{figure}

\begin{figure}
    \centering
    \includegraphics[width=1\linewidth]{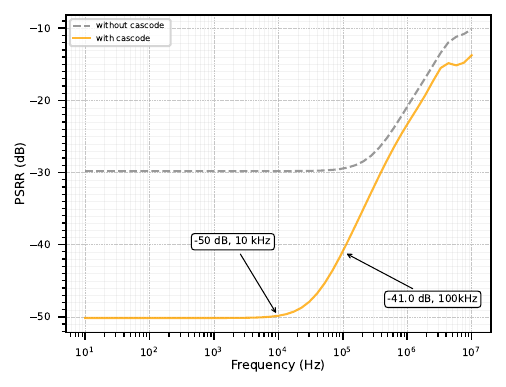}
    \caption{Power Supply Rejection (PSRR) of the voltage reference (with and without cascode)}
    \label{fig:psrr}
\end{figure}

The voltage reference proposed in this paper has been implemented on a 90 nm CMOS process with the circuit schematic given in Fig. \ref{fig:full-schematic}. The circuit has been optimized to give an excellent temperature coefficient while being power efficient, with a Quiescent current of 1.3 \(\mu\)A at room temperature. The power consumption ranges from 0.67 \(\mu\)W with a supply of 0.5V to 4.3\(\mu\)W with a supply of 3.3V. The different components of the output current generated, i.e., the CTAT, PTAT, and the curvature compensation current, can be seen in Fig. \ref{fig:all-currents}. CTAT current has a slope of -0.549 nA/\textdegree C, PTAT has a slope of 0.731 nA/\textdegree C, and the curvature compensation current has a best fit slope of 0.194 nA/\textdegree C. The slope of the net current \(I_{REF}\) flowing through \(R_3\) can be calculated as
\begin{figure}[b!]
    \centering
    \includegraphics[width=1.03\linewidth]{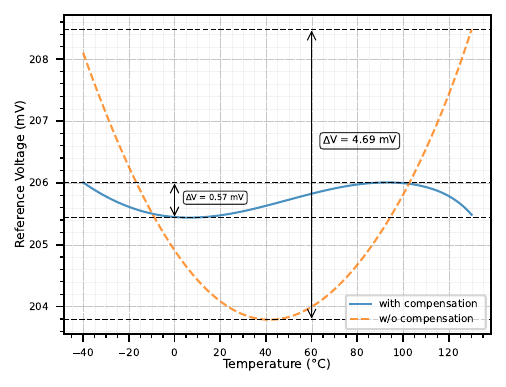}
    \caption{Variation of output reference voltage (\(V_{REF})\) over a wide temperature from -40\textdegree C to 130\textdegree C}
    \label{fig:vref-temp}
\end{figure}
\[I_{REF} = I_{ptat}' + I_{ctat}'-I_{comp}\]
\begin{table*}[!b]
\centering
\renewcommand{\arraystretch}{1.35}
\caption{Performance comparison with prior sub-1V CMOS references.}

\begin{tabular}{|
>{\centering\arraybackslash}m{4.3cm}|
>{\centering\arraybackslash}m{2.0cm}|
>{\centering\arraybackslash}m{2.4cm}|
>{\centering\arraybackslash}m{2.3cm}|
>{\centering\arraybackslash}m{1.3cm}|
>{\centering\arraybackslash}m{2.4cm}|}
\hline
\textbf{Parameter} & \textbf{Ref.~\cite{ADC}} & \textbf{Ref.~\cite{add1}} & \textbf{Ref.~\cite{add2}} & \textbf{Ref.~\cite{add3}} & \textbf{This Work} \\ \hline
CMOS Process (nm) & 65 & 180 & 180 & 180 & \textbf{90} \\ \hline
Min. Supply Voltage (V) & 0.7 & 0.8 & 0.95 & 0.6 & \textbf{0.5} \\ \hline
Supply Range (V) & 0.7–3 & 0.8–2 & 0.95–2.2 & 0.6–1 & \textbf{0.5–3.3} \\ \hline
Ref. Voltage (mV) & 223 & 332 & 773 & 404 & \textbf{205} \\ \hline
Power Consumption (nW) & 75 & 15.6 & 5.6 & 840 & \textbf{670} \\ \hline
Line Sensitivity (\%/V) & 2.54 & 0.5 & 0.073 & 1.8 & \textbf{1.65} \\ \hline
PSRR (dB) & --62 dB @1 kHz & --44.7 dB @100 Hz & --41 dB @100 Hz & -- & \textbf{--50 dB @10 kHz} \\ \hline
Temp. Range (°C) & --60–160 & 0–120 & --20–80 & --40–125 & \textbf{--40–130} \\ \hline
Temp. Coefficient (ppm/°C) & 19.47 & 62.8 & 88 & 40 & \textbf{16.28} \\ \hline
\end{tabular}
\end{table*}
\begin{figure*}[!ht]
    \centering
    \begin{subfigure}{0.49\textwidth}
        \centering
        \includegraphics[width=\linewidth]{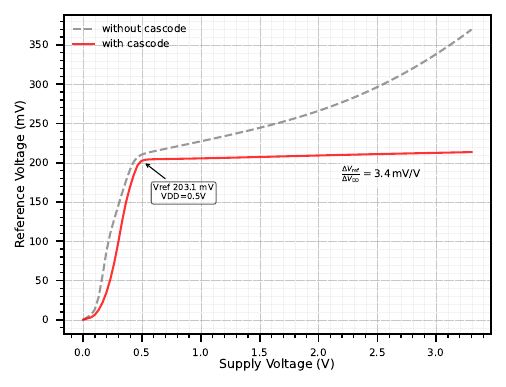}
        \caption{}
        \label{fig:line-sensitivity}
    \end{subfigure}
    \hfill
    \begin{subfigure}{0.49\textwidth}
        \centering
        \includegraphics[width=\linewidth]{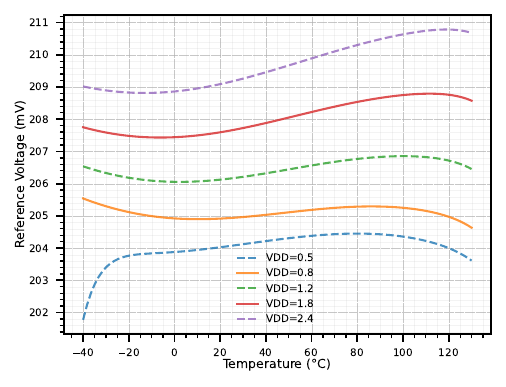}
        \caption{}
        \label{fig:tc-vdd}
    \end{subfigure}
    \caption{\(V_{REF}\) vs supply and temperature variation (a) Output reference voltage \(V_{REF}\) as a function of supply voltage (with and without cascode) (b) Temperature variation of \(V_{REF}\) at different supply voltages}
    \label{fig:combined}
\end{figure*}
\[\text{Slope}(I_{REF}) = \text{11 pA/°C} \approx\ \text{0}\]
Fig. \ref{fig:vref-temp} shows the variation of output voltage \(V_{REF}\) versus a wide range temperature of -40\textdegree C to 130\textdegree C with and without curvature compensation. Without compensation, the circuit has a temperature coefficient of 137 ppm/\textdegree C with a maximum deviation of 4.69 mV. However, with compensation, the proposed circuit achieves a brilliant temperature coefficient of 16.28 ppm/\textdegree C with a maximum deviation of just 0.57 mV or 570 \(\mu\)V. The reference voltage produced at room temperature (27\textdegree C) is 205.5 mV, with a minimum \(V_{REF}\) of 205.44 mV at 8\textdegree C and a maximum \(V_{REF}\) of 206.01 mV at 92\textdegree C. Fig. \ref{fig:line-sensitivity} shows the voltage supply characteristics of the voltage reference with and without cascode current mirror modification. It can be seen that the modification has significantly lowered the voltage supply sensitivity to having a line sensitivity of just 1.65\%/V, which is only about a 3.4 mV increase in \(V_{REF}\) voltage for each volt of increase in supply voltage which can also be seen in Fig. \ref{fig:tc-vdd}. The voltage reference reaches its nominal value of 205 mV (with less than 1\% error) for a tremendously low minimum supply voltage of 0.5 V. The voltage reference nominal supply voltage is 1V and has a wide range of operational voltage starting from 0.5 V to over 3.3 V. Fig. \ref{fig:psrr} shows the plot of power supply rejection of the proposed voltage reference with and without cascode current mirror modification. From a PSRR of -30 dB, the voltage reference achieves a PSRR of -50 dB up to a frequency of 10 kHz with the modification. The circuit also achieves a PSRR of -41 dB at a high frequency of 100 kHz.


\clearpage
\balance 
\section{CONCLUSION}
This paper presents a fully-integrated CMOS voltage reference using low voltage threshold transistors in a 90 nm process node. The reference voltage attains independence of temperature variation by the mutual compensation of PTAT, CTAT, and curvature compensation currents. By exploiting subthreshold and weak inversion characteristics of MOSFETs, the design achieves a very low temperature coefficient, line sensitivity, and power consumption. Monte-carlo simulations show a reference voltage of 204.85 mV with a temperature coefficient of 16.28 ppm/\textdegree C in wide range of -40 \textdegree C to 130 \textdegree C, consuming just 0.67 \(\mu\)W at a 0.5 V supply and 1.3 \(\mu\)W at a 1 V supply. The performance summary and comparison with recent works are given in Table 1.
\section*{Acknowledgment}
The authors would like to thank the Chips to Startup (C2S) program, Ministry of Electronics and Information Technology (MeitY), Government of India, for providing access to the necessary EDA tools and resources that enabled this work. We also gratefully acknowledge Mr. Buddhi Prakash Sharma, Ph.D. scholar at Birla Institute of Technology and Science, for his valuable assistance with Monte Carlo simulations.

\vspace{12pt}

\end{document}